# Encryption Device Based on Wave-Chaos for Enhanced Physical Security of Wireless Wave Transmission

Hong Soo Park and Sun K. Hong
*School of Electronic Engineering, Seoul, 06978, South Korea*

We introduce an encryption device based on wave-chaos to enhance the physical security of wireless wave transmission. The proposed encryption device is composed of a compact quasi-2D disordered cavity, where transmit signals pass through to be distorted in time before transmission. On the receiving end, the signals can only be decrypted when they pass through an identical cavity. In the absence of a proper decryption device, the signals cannot be properly decrypted. If a cavity with a different shape is used on the receiving end, vastly different wave dynamics will prevent the signals from being decrypted, causing them to appear as noise. We experimentally demonstrate the proposed concept in an apparatus representing a wireless link.

## I. INTRODUCTION

Wave-chaotic systems exhibit extreme sensitivity to small changes in their configuration or initial wave conditions in the short wavelength regime [1]. This intriguing property of wave-chaotic systems has led to research activities in various areas including statistical electromagnetics [2-12], wireless communications [13,14], wave-front shaping [15,16], sensors [17-19], high-power electromagnetics [20-24], acoustic, microwave and optical cavities [25-31], and quantum dots [32,33].

A disordered cavity with electrically large dimensions is a wave-chaotic system that can be utilized with electromagnetic waves. For a two-port disordered cavity, a slight change in its shape or position of the input and/or output ports would result in vastly different ray trajectories between the ports, producing uncorrelated transfer functions (impulse responses) before and after the change [1]. Due to such sensitivity of a wave chaotic system, it is very difficult to replicate the system's response without the knowledge of the system to a high degree of accuracy.

Here we introduce a new application of the electromagnetic wave-chaotic system. That is, an encryption device that can be utilized to enhance the physical security of wireless electromagnetic wave transmission. The proposed device is made up of an electrically large but physically compact quasi-2D disordered cavity [34]. A signal passes through the cavity before transmission, which is then distorted in a random fashion. Such a distortion acts as physical encryption of the signal. The encrypted signal is then radiated through a transmitting antenna. On the receiving end, the signal can only be decrypted when it passes through a device identical to that used for encryption. If a device (cavity) other than the one used for encryption is used (or no device is used at all), completely different wave dynamics in that device will prevent the signal from being decrypted.

While the proposed concept can be applied to a range of frequency regimes and signal types, here we demonstrate the concept by applying it to ultra-wideband (UWB) signals in the microwave regime. In this process, we utilize three different wave-chaotic cavities with uncorrelated impulse responses, which are used to encrypt UWB pulses carrying digital information. Through measurement, we show that the transmitted signals can be correctly received and decrypted only when identical devices are used on both the transmitting and receiving ends. A key advantage of the proposed concept is in the simplicity of implementation, where encryption is done by simply letting a signal pass through a compact wave-chaotic cavity.

## II. ENCRYPTION DEVICE AND SYSTEM BASED ON WAVE CHAOS

Fig. 1 illustrates a notional system employing the proposed encryption concept under the context of wireless communications using binary phase-shift keying (BPSK) type modulation. The proposed system consists of the following processes. In the transmitter (Alice), a short pulse $\delta(t)$ is generated and split into two paths by a switch having a period of $T$. The first path allows $\delta(t)$ to pass unaltered, while the second path modulates $\delta(t)$ with digital information (digital symbol '10' used in Fig. 1) and passes it through the encryption device (wave-chaotic cavity), which encrypts the pulse by producing impulse response $h(t)$ or $-h(t)$ based on the symbol ($h(t)$ for '1' and $-h(t)$ for '0'). The split paths are then combined to produce an encrypted signal $x(t)$, which is a chain of signal pairs containing $\delta(t)$ and an encrypted symbol $h(t)$ or $-h(t)$ in alternate periods. The encrypted signal $x(t)$ is consequently transmitted through an antenna.

---

*Corresponding author: shong215@ssu.ac.kr

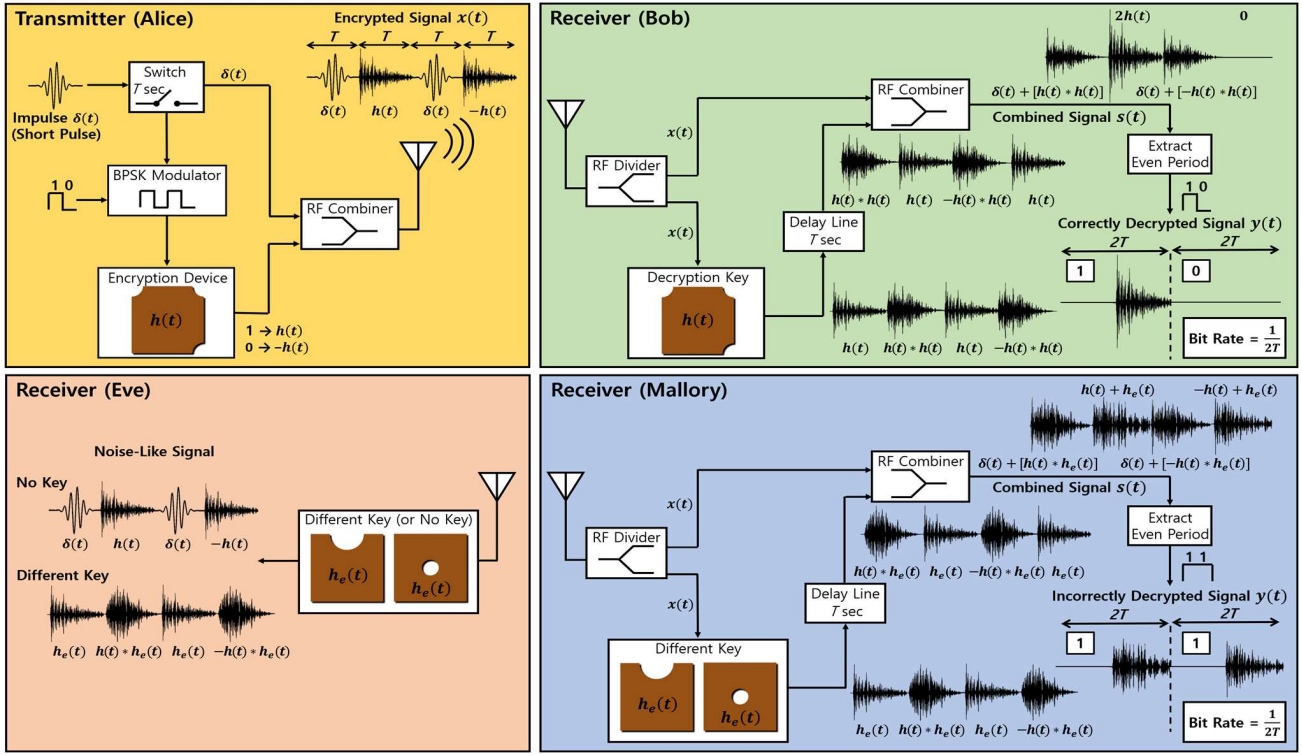

FIG. 1. A schematic of the proposed concept of encryption and decryption process based on wave-chaos. Given a notional transmit system (Alice), three different possible receivers are illustrated: a legitimate receiver (Bob), a passive eavesdropper (Eve), and a malicious receiver (Mallory).

On the receiving end, we consider three different types of receivers, namely Bob, Eve, and Mallory. For a legitimate receiver (Bob), the receiving structure includes a device identical to that used on the transmitting end for decryption of signals. The received signal $x(t)$ is simultaneously split into two paths through a divider, where the first path lets $x(t)$ pass unaltered, while the second path contains a delay line and the same wave-chaotic cavity used in the transmitter as a decryption key. As $x(t)$ passes through the cavity in the second path, the cavity output signal becomes a train of signal pairs consisting of the impulse response $h(t)$ of the decryption key and $h(t)*h(t)$ or $-h(t)*h(t)$ in alternate periods, where $*$ denotes convolution. The cavity output signal is then time-shifted by $T$ with a delay line, which aligns $h(t)$ (generated from the decryption key) with $h(t)$ or $-h(t)$ (encrypted pulse from encryption device) when the two paths are combined. Note that using the same cavities for encryption and decryption, $h(t)$ of the second path contains the exact same distortion profile as in the encrypted $h(t)$ or $-h(t)$ of the first path. After extracting the even periods in the combined signal $s(t)$, i.e. filtering out every other $T$ containing $\delta(t)$ and $h(t)*h(t)$ or $-h(t)*h(t)$, the resulting signal $y(t)$ only consists of either $2h(t)$ or $0$ (null), which respectively represents the original information '1' or '0' with a symbol period of $2T$ (bit rate of $1/2T$). As a result, digital information would be received correctly when the same device as that used in the transmitter is utilized as a decryption key to decode the encrypted signal.

There may also exist eavesdroppers who attempt to intercept the information, and they could be classified as passive and malicious eavesdroppers. For a passive eavesdropper (Eve), the overall structure of the receiver for the decryption process (e.g. two split paths, delay line, even period extraction, etc.), as well as the encryption device, is unknown. Accordingly, Eve may use a device different from that of the transmitter as a receiving cavity (or no cavity) for eavesdropping. Due to the absence of the decryption process, when a cavity is not used, the encrypted signal $x(t)$ is received unchanged, and when a cavity different from the encryption device is used, the signal corresponding to the convolution of

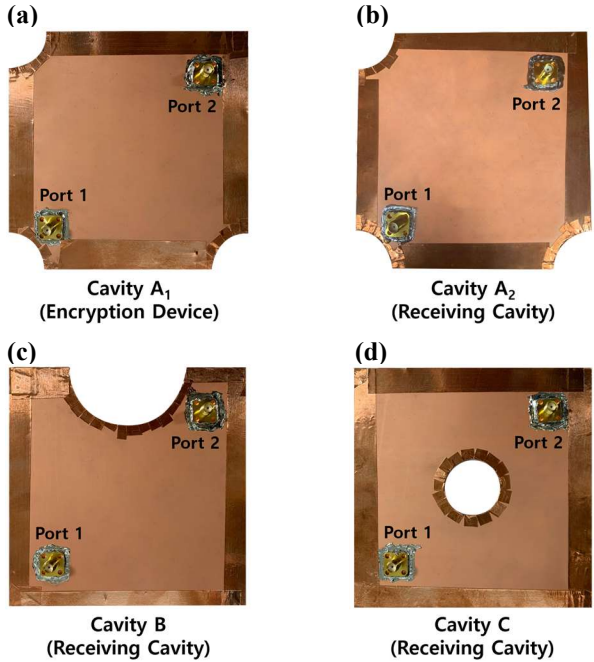

FIG. 2. Fabricated three different wave-chaotic cavities used in experiment: (a) Cavity $A_1$ for encryption on the transmit side (Alice). (b) Cavity $A_2$ for legitimate decryption (Bob). (c) Cavity B and (d) Cavity C for decryption in eavesdropping (Mallory).

$x(t)$ and the impulse response of the receiving cavity $h_e(t)$, i.e. $x(t) * h_e(t)$, is received. In both cases, the received signal appears to be noise, and therefore the information would not be intercepted by eavesdroppers. On the other hand, for a malicious eavesdropper (Mallory) who can mimic the receiving structure but uses a different cavity than that used for encryption, the decryption process is also applied to $x(t)$ in the same way as Bob. Despite the use of the same decryption process, $h_e(t)$ exhibiting a different distortion profile from the encrypted $h(t)$ or $-h(t)$ cannot recover the original information via producing $2h(t)$ or $0$, and the resulting signal will contain different digital information than originally encoded. That is, a slight difference from the encryption device results in vastly different wave dynamics, which prevents any other cavities from being a proper decryption key. In other words, such a decryption process would be only possible when the receiver uses an identical cavity used to encrypt the signal, meaning that the wave-chaotic cavity can be simply utilized as a unique encryption device for physical security.

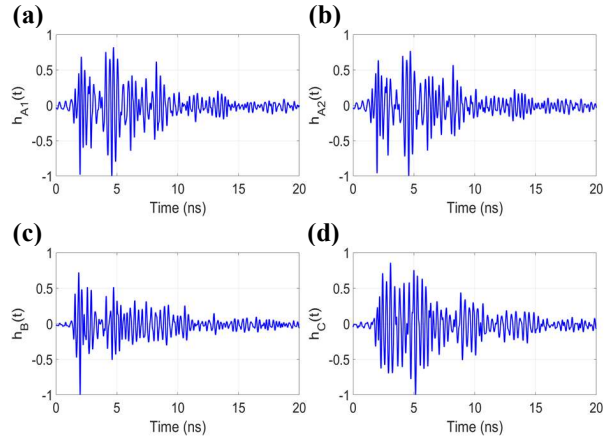

FIG. 3. Measured impulse responses of each cavity: (a) $h_{A1}(t)$ of Cavity $A_1$. (b) $h_{A2}(t)$ of Cavity $A_2$. (c) $h_B(t)$ of Cavity B. (d) $h_C(t)$ of Cavity C.

TABLE I. Correlation coefficient between each cavity

| Encryption Device | Receiving Cavity | Correlation Coefficient |
|---|---|---|
| Cavity $A_1$ | Cavity $A_2$ | 0.9413 (high) |
| | Cavity B | 0.4486 (moderate) |
| | Cavity C | 0.143 (low) |

## III. EXPERIMENTS

The proposed concept is validated via measurement using compact two-port quasi-2D wave-chaotic cavities. Fig. 2 shows three different cavities, namely Cavities A, B, and C, designed on a 3.175 mm-thick RT/duroid 6010 series laminate with broadside dimensions of $12.7 \times 12.7$ mm$^2$. In the case of Cavity A, two identical versions, namely Cavity $A_1$ and Cavity $A_2$, are fabricated to be utilized for encryption and decryption, respectively. The sidewalls of all cavities are sealed with copper tape. As shown in the figure, three corners of Cavity A, the top of Cavity B, and the center of Cavity C are cut circularly to provide different shapes and thus different wave dynamics. Although all cavities share the same footprint, the three different shapes lead to distinct distortion profiles in each cavity, resulting in mutually uncorrelated impulse responses from one another.

We consider that Cavity $A_1$ is used as an encryption device on the transmit side and then compare the results when Cavity $A_2$ (legitimate scenario), Cavities B and C (eavesdropping scenarios) are used as receiving cavities on the receiving end. Also, it is assumed that the decryption process is correctly applied in the receiving structure using each cavity (which means that the legitimate scenario is Alice to Bob, and the eavesdropping scenario is Alice to Mallory). In particular, to confirm the decryption of the signal according to the degree of correlation, the cavities are designed to have impulse responses with different correlations, i.e. high, moderate and low correlations. The measurement of the impulse response (between Port 1 and Port 2) is carried out in the time domain using an arbitrary waveform generator (Tektronix AWG7102) and oscilloscope (Tektronix TDS6154C). Fig. 3 shows the measured impulse response of each cavity, i.e. $h_{A1}(t)$, $h_{A2}(t)$, $h_B(t)$, and $h_C(t)$, when a short UWB (bandwidth of 3 – 10 GHz) pulse is used as the input signal, and the correlation coefficients between each cavity pair are shown in Table 1. It can be seen that the correlation with the impulse response of the encryption device (Cavity $A_1$) varies depending on the shape of the receiving cavity. That is, Cavity $A_2$ having the same shape as the encryption device has a high correlation (close to 1) as expected, whereas Cavities B and C with different shapes from the encryption device have moderate (close to 0.5) and low correlations (close to 0), respectively.

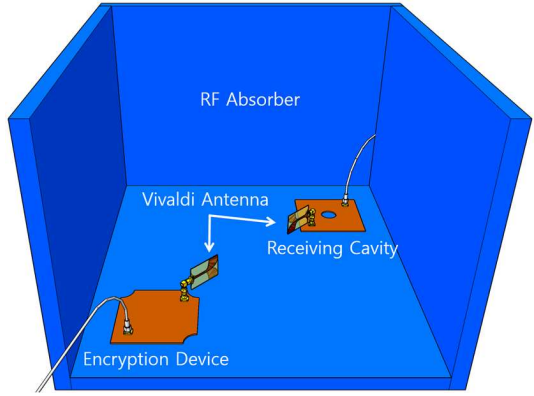

FIG. 4. An illustration of the experiment setup to measure channel responses between the encryption device (Cavity $A_1$) and receiving cavities (Cavities $A_2$, B, and C)

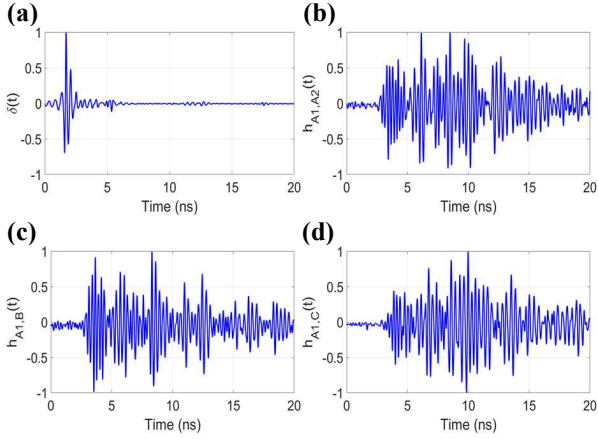

FIG. 5. Measured antenna response and channel responses between encryption device (Cavity $A_1$) and receiving cavities (Cavities $A_2$, B, and C): (a) antenna response $\delta(t)$. (b) channel response $h_{A1,A2}(t)$ using Cavity $A_2$. (c) channel response $h_{A1,B}(t)$ using Cavity B. (d) channel response $h_{A1,C}(t)$ using Cavity C.

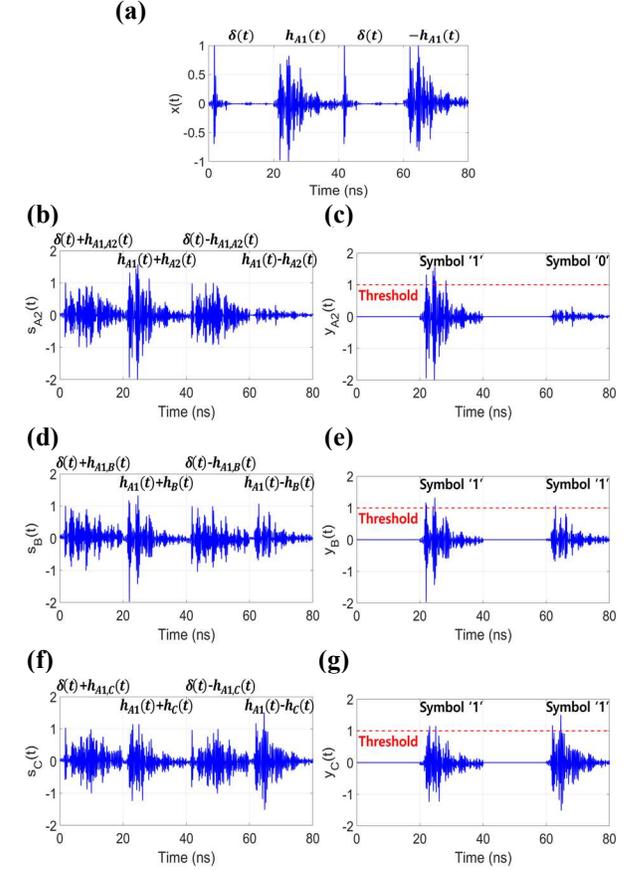

FIG. 6. The encrypted signal and the output signals in the receiver when Cavity $A_1$ is used as the encryption device for three cases. (a) the encrypted signal $x(t)$. (b) the combined signal $s_{A2}(t)$ using Cavity $A_2$. (c) output signal $y_{A2}(t)$ where even period is extracted from $s_{A2}(t)$. (d) the combined signal $s_B(t)$ using Cavity B. (e) output signal $y_B(t)$ where even period is extracted from $s_B(t)$. (f) the combined signal $s_C(t)$ using Cavity C. (g) output signal $y_C(t)$ where even period is extracted from $s_C(t)$.

Now, to obtain the $h(t)*h(t)$ or $h(t)*h_e(t)$ part of the received signal (see Fig. 1), the channel responses between the transmitter and receiver (one encryption device and receiving cavity pair) are measured in an apparatus as shown in Fig. 4. Similar to the impulse response measurements, the measurement is performed in the time domain using an UWB pulse as the input signal, and the transmitting and receiving antennas (each connected to a cavity) are placed about 50 cm from each other, which are surrounded by RF absorbers to minimize unwanted reflection and noise. Here we use Vivaldi antennas designed to cover 3 – 10 GHz on both ends. The antenna response without cavities is also measured to obtain the $\delta(t)$ part of the received signal. Fig. 5 shows the antenna response (no cavities connected) and channel responses (cavities connected) for three different pairs (i.e. $A_1$ to $A_2$, $A_1$ to B, and $A_1$ to C). Note that the antenna response means $\delta(t)$ of the received signal $x(t)$, and the channel responses $h_{A1,A2}(t)$, $h_{A1,B}(t)$, and $h_{A1,C}(t)$ correspond to $h_{A1}(t)*h_{A2}(t)$, $h_{A1}(t)*h_B(t)$, and $h_{A1}(t)*h_C(t)$, respectively.

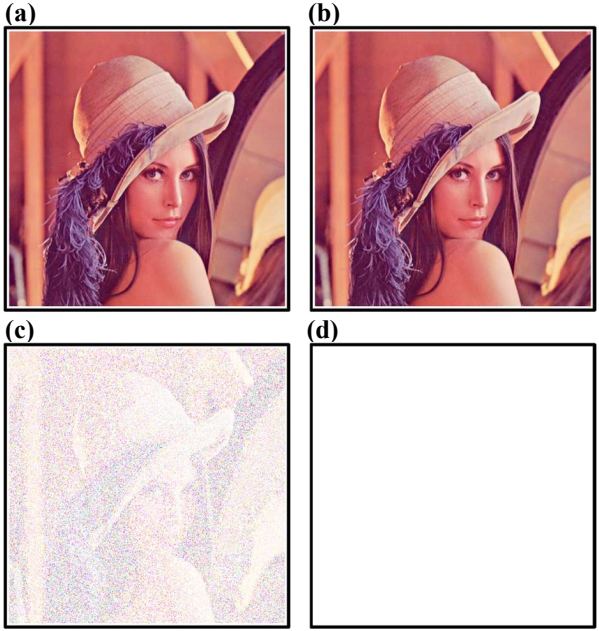

FIG. 7. (a) Original image. (b) Decrypted image using Cavity $A_2$. (c) Decrypted image using Cavity B. (d) Decrypted image using Cavity C.

## IV. RESULTS AND DISCUSSION

Using the measured signals, decryption is done through signal processing and the results from three cavities are compared. In Fig. 6, the encrypted signal $x(t)$ from the encryption device (Cavity $A_1$), and the resulting signals $s(t)$ and $y(t)$ by using the aforementioned cavities as a receiving cavity are plotted. It is assumed that digital information of '10' is transmitted, and the period $T$ is set to 20 ns. Note that since $T$ depends on the decay time of the cavity, the symbol period can be reduced by changing the electrical size of the cavity. As described in Fig. 1, it can be seen that $x(t)$ contains $\delta(t)$ and encrypted signals $h_{A1}(t)$ and $-h_{A1}(t)$ (corresponding to symbols '1' and '0') in alternate periods (Fig. 6(a)). Also, $s(t)$ represents the sum of the first and second path signals of the receiver, and $y(t)$ represents the resulting signal obtained by extracting the even period of $s(t)$.

For Cavity $A_2$ having a high correlation with the encryption device (legitimate scenario), the combined signal $s_{A2}(t)$ consists of $\delta(t) \pm h_{A1,A2}(t)$, $h_{A1}(t)+h_{A2}(t)$, and $h_{A1}(t)-h_{A2}(t)$ (Fig. 6(b)). Therefore, $h_{A1}(t)+h_{A2}(t)$ and $h_{A1}(t)-h_{A2}(t)$ become $2h_{A1}(t)$ (or $2h_{A2}(t)$) and 0 (null), respectively, with high degree of accuracy. The resulting signal $y_{A2}(t)$ as a result of even period extraction remains only the signals corresponding to $2h_{A1}(t)$ and null (Fig. 6(c)), which can then be converted into the original information '10' with the symbol period of 40 ns by applying an appropriate threshold level. That is, the signal $2h_{A1}(t)$ with a magnitude greater than the threshold would be a symbol '1', and the null signal with a magnitude smaller than the threshold would be a symbol '0'.

On the contrary, for Cavities B and C, which respectively exhibit moderate and low correlations with the encryption device (eavesdropping scenario), $h_{A1}(t) \pm h_B(t)$ and $h_{A1}(t) \pm h_C(t)$ parts of $s_B(t)$ and $s_C(t)$ do not result in $2h_{A1}(t)$ and null (Figs. 6(d) and 6(f)). Hence, despite the thresholding process after extracting the even period of the signals $s_B(t)$ and $s_C(t)$, both $y_B(t)$ and $y_C(t)$ produce an incorrect symbol of '11' rather than the original digital information '10' (Figs. 6(e) and 6(g)), indicating that the information is not properly decrypted. This implies that the null signal from the difference in impulse responses with high correlation plays a significant role in decoding the signal, and the absence of the null signal due to low correlation makes the original information unrecoverable.

We also demonstrate the proposed encryption process using the 'Lena' image with $512 \times 512$ pixels. Here the image is separated into RGB components, and each pixel converted to an 8-bit binary number is encrypted through Cavity $A_1$. After the encrypted symbols are decoded through each receiving cavity, the RGB components are recombined to restore the image. The original image and the resulting images from the decoding process are shown in Fig. 7. The image decoded with Cavity $A_2$ (Fig. 7(b)) shows a perfect reproduction of the original image, indicating that Cavity $A_2$, which has a high correlation with the encryption device (Cavity $A_1$), successfully operates as a decryption key. However, the images decoded with Cavities B and C are not properly recovered. In Fig. 7(c), a recovered image from signals that went through Cavity B is shown, which loosely reconstructs the original image, but with very low fidelity indicating that the information is not properly restored. In Fig. 7(d), a recovered image from signals through Cavity C is shown, where only white pixels are shown, indicating that the information is completely lost.

These results demonstrate that the proposed encryption scheme using wave-chaotic cavities can be utilized to enhance the physical security of wave transmission in that it only allows decryption of the signal when a cavity highly correlated with the encryption device is used as a receiving cavity. Since a small change in the shape of a two-port disordered cavity leads to vastly different ray trajectories between the ports, the original information cannot be restored unless the cavity is identical to that used for encryption, which provides a highly enhanced additional layer of physical security.

## V. CONCLUSIONS

In this paper, we proposed and validated a novel encryption scheme to enhance the physical security of wireless wave transmission leveraging wave-chaos. By using a compact two-port wave-chaotic cavity, it was experimentally shown that the distortion that occurs in a random fashion as the signal passes through the cavity itself serves as physical encryption of the signals. The overall results demonstrate that the encrypted signals can be correctly decoded only when identical devices are used in both the transmitter and receiver, while the absence of an identical device in the receiver prevents the signals from being properly decrypted. The proposed concept can be utilized in applications such as wireless communications and radars, and further be extended to a variety of frequency regimes, signal types, and modulations.